\documentclass[%
 reprint,
 amsmath,amssymb,
 aps,
prl,
]{revtex4-2}

\usepackage{graphicx}
\usepackage{dcolumn}
\usepackage{bm}
\usepackage{hyperref}

\usepackage{amsmath}

\usepackage{xcolor}

\begin{document}

\title{An Improved Fit for Linear Halo Bias at High Redshift}

\author{{Kuan} {Wang}}
\email{kuan.wang@austin.utexas.edu}
\affiliation{University of Texas at Austin, Department of Astronomy, 2515 Speedway, Austin, TX 78712, USA}
\thanks{ORCID: \href{https://orcid.org/0000-0001-7690-2260}{0000-0001-7690-2260}}

\author{{Julian B.} {Mu\~noz}}
\affiliation{University of Texas at Austin, Department of Astronomy, 2515 Speedway, Austin, TX 78712, USA}
\affiliation{Cosmic Frontier Center, Austin, TX 78712, USA}
\affiliation{Texas Center for Cosmology \& Astroparticle Physics, Austin, TX 78712, USA}
\thanks{ORCID: \href{https://orcid.org/0000-0002-8984-0465}{0000-0002-8984-0465}}

\author{{L. Y. Aaron} {Yung}} 
\affiliation{Space Telescope Science Institute, 3700 San Martin Drive, Baltimore, MD 21218, USA}
\thanks{ORCID: \href{https://orcid.org/0000-0003-3466-035X}{0000-0003-3466-035X}}

\date{\today}

\begin{abstract}
High- to ultrahigh-redshift clustering of halos provides a powerful tool to understand cosmology and galaxy formation.
However, theoretical predictions are not firmly established in the first billion years, where current and upcoming surveys are beginning to reach percent-level precision. 
Here we measure dark matter halo biases at $z=6$ - 19 from simulation data, and find they are $\sim$ 3 - 4$\%$ higher than canonical results calibrated at low $z$.
We provide an updated linear-bias fit at these early times, reducing the mean systematic offset to $< 1\%$.
These results will enable robust interpretation of early-Universe galaxy clustering from JWST, Roman, and intensity-mapping surveys.
\end{abstract}

\maketitle

\section{Introduction}
\label{sec:intro}

In the standard cosmological paradigm, galaxies reside within dark matter halos \cite{whiterees78,blumenthal84}, which act as the key link between theoretical predictions (e.g., of the cosmological matter field) and observations (of galaxies, clusters, and and other large-scale structure tracers).
Accurate modeling of halo behavior is therefore essential for interpreting survey measurements, constraining cosmological models, and understanding galaxy formation \cite{cooray-sheth02,zheng2005,zheng2007,wechsler-tinker2018,behroozi2019,yung22,yung23}.

A simple yet powerful way to study halos is through their {\it bias}, which quantifies how the spatial distribution of halos relates to that of the dark matter \cite{kaiser84,mo-white96}.
Over the past several decades, halo bias has been studied extensively at low redshift ($z\lesssim 4$) using both analytic approaches and numerical simulations.
Early theoretical developments based on the peak–background split framework \cite{mo-white96,sheth-tormen99} have been followed by increasingly precise calibrations using large suites of $N$-body simulations.
These efforts have converged in fitting functions expressing halo bias as a function of halo mass or peak height \cite[e.g.,][]{jing98,smt2001,tinker2010,bhattacharya2011}, which are widely used in galaxy clustering analyses.

In contrast, halo bias at high redshift ($z\gtrsim 4$) remains comparatively underexplored, barring a small number of studies \cite[e.g.,][]{cohn08,jose2016,nasirudin2020,wang22}, which yielded diverse results.
Most existing fitting functions are calibrated using simulations at low redshift and are often extrapolated to early times without direct validation (see, e.g.,~\cite{reed2009,mirocha2021,shuntov2025}).
However, high-redshift halos form in a less evolved, more linear density field, where clustering behavior may differ from their low-redshift counterparts \cite[e.g.,][]{kravtsov12,desjacques18}.
As a result, the accuracy and applicability of low-redshift bias calibrations at high redshift remain open questions.

This issue has become increasingly important as observations begin to probe the high-redshift universe with unprecedented depth.
Current and upcoming facilities, including the James Webb and  Roman Space telescopes, as well as intensity-mapping experiments like the Hydrogen Epoch of Reionization Array (HERA) or the Square Kilometer Array (SKA)~\cite{JWST2006,Roman2015,HERA2017,SKA2015}, are delivering meaningful measurements of clustering at high redshift with various types of tracers.
While these surveys promise powerful constraints on structure formation at early times, they also highlight the need for accurate theoretical models of halo bias in this regime.

In this {\it Letter} we calibrate a new fit for the linear halo bias at high redshifts.
We use the recent \textsc{GUREFT} suite of simulations \cite{gureft}, which combines four different-resolution boxes specifically designed for high- to ultrahigh-$z$ analyses, and use halo catalogs constructed with \texttt{ROCKSTAR} \cite{ROCKSTAR}, which takes into account the full six-dimensional phase-space information to robustly track structures in the dense and rapidly evolving high-redshift density field.
Our main result in \autoref{eq:biasfit_us} provides an update of the canonical \citet{tinker2010} bias fit, improving the mean agreement with simulations to $<1\%$ for halos at $z = 6-19$.
Throughout this work we assume a flat $\Lambda {\rm CDM}$ cosmology with $\Omega_m = 0.307, \Omega_\Lambda = 0.693, H_0 = 67.8\, {\rm km}\ {\rm s}^{-1} {\rm Mpc}^{-1}, \sigma_8 = 0.829$, and $n_s = 0.960$ \cite{planck16}.
Unless otherwise specified, all distances are expressed in comoving ${\rm Mpc}$, and all masses in units of $M_\odot$.

\section{Methods}
\label{sec:methods}

\subsection{Simulation suite and samples}
\label{sec:sim_sample}

The \textsc{GUREFT} simulation suite \cite{gureft} consists of four periodic boxes with comoving side lengths of 5, 15, 35, and 90 $h^{-1}{\rm Mpc}$, each containing $1024^3$ particles.
We list the corresponding particle mass resolutions in \autoref{tab:spec_box}. 
The smallest box has a particle mass of $1.46\times10^4M_\odot$, which allows us to resolve low-mass halos down to a few times $10^6M_\odot$.
These halos correspond to more common peaks in the density field and are particularly relevant at high redshift.
The strategically chosen dynamic range of the \textsc{GUREFT} simulation boxes, together with their high mass resolution, makes them especially well suited for studying the population and clustering of high-redshift halos.
We refer the interested reader to Ref.~\cite{gureft} for further details.

\renewcommand{\arraystretch}{1.3}
\begin{table*}[]
    \centering
    \begin{ruledtabular}
    \begin{tabular}{ c c c c c c }
         Box & $L_{\rm box}\left[{h^{-1}\rm Mpc}\right]$ & $m_{\rm ptcl}\left[M_\odot\right]$ & $z$ & $\log_{10} M_{\rm low}\left[M_\odot\right]$ & $\log_{10} r\left[{\rm Mpc}\right]$ \\
         \hline\hline
         GUREFT-05 & 5 & $1.46\times 10^4$ & $[6, 19]$ & 6.5, 7.0, 7.5, 8.0 & $[-1.8, 0.5]$ \\
         GUREFT-15 & 15 & $3.95\times 10^5$ & $[6, 15]$ & 8.0, 8.5, 9.0, 9.5 & $[-1.3, 1.0]$ \\
         GUREFT-35 & 35 & $5.02\times 10^6$ & $[6, 12]$ & 9.0, 9.5, 10.0 & $[-1.0, 1.4]$ \\
         GUREFT-90 & 90 & $8.53\times 10^7$ & $[6, 10]$ & 10.0, 10.5, 11.0 & $[-0.6, 1.8]$ 
    \end{tabular}
    \end{ruledtabular}
    \caption{Simulation box properties and analysis ranges.
    For each simulation box, we list the particle mass $m_{\rm ptcl}$, the redshift range that we analyze, the lower bounds of mass $\log_{10} M_{\rm low}$ for sample selection, and the scale range $\log_{10} r$ used in the clustering measurements.}
    \label{tab:spec_box}
\end{table*}

Halo catalogs are constructed using the \texttt{ROCKSTAR} halo finder \cite{ROCKSTAR}.
\texttt{ROCKSTAR} halo finding is based on adaptive hierarchical refinement of friends-of-friends groups in six phase-space dimensions and one time dimension.
By incorporating velocity information rather than relying solely on spatial overdensity, \texttt{ROCKSTAR} robustly separates nearby structures and accurately identifies bound halos, even in dense or rapidly evolving environments.
Such conditions are common at high redshift, where halos are less virialized and mergers occur frequently \cite{stewart09,wang20}.
As a result, \texttt{ROCKSTAR} provides stable halo properties and consistent catalogs across snapshots, making it particularly well suited for constructing reliable halo samples for high-redshift studies. 
We exclude subhalos from the catalogs, and restrict our analysis to host halos, i.e., the halos that do not reside within the virial radius of a more massive system.
We note that backsplash halos (apparently isolated halos that used to be subhalos \cite{diemer2021}) are treated as hosts under this criterion.
We consider snapshots at integer redshifts.

For the halo bias estimation, we define mass-bin samples of width 0.5 dex at integer and half-integer values of $\log_{10}(M_{\rm vir}/M_\odot)$. 
We require a minimum sample size of 1000 halos, which results in different redshift ranges and mass bins for the different boxes; we list them in \autoref{tab:spec_box}.
We have tested that using mass-threshold samples instead of mass-bin samples yield consistent results.
We also use the positions of dark matter particles in the simulation to compute halo–matter correlation functions for the bias estimation.
To reduce computational cost, we randomly downsample the particle catalogs used for the correlation functions by a factor of 1000.
We have verified that the effect of this downsampling on our measurement is insignificant.

\subsection{Correlation function measurement}
\label{sec:cf_method}

We calculate halo bias through correlation functions $\xi_{ij}$, where $i,j\in \{h,m\}$ label halos and matter; here $\xi_{hh}$ is the halo auto-correlation, $\xi_{hm}$ the cross-correlation with matter, and $\xi_{mm}$ the matter auto-correlation.
For the measurement of correlation functions, we adopt the Landy-Szalay \cite{landy-szalay93} estimator based on pair counting, and perform the measurement with the \texttt{pycorr} \cite{corrfunc,Sinha2020} package.
We measure correlation functions for logarithmic bins of scales, whose edges are set at integer multiples of 0.1 dex.
These bins are truncated at a minimum scale of three times the virial radius ($3R_{\rm vir}$) of the lowest-mass halos considered in each box, and at a maximum scale of half the box size.
The respective ranges of scales for each box are listed in \autoref{tab:spec_box}.
We will further narrow down the range of scales appropriate to each sample in the analysis.

\begin{figure*}
    \centering
    \includegraphics[width=0.8\linewidth]{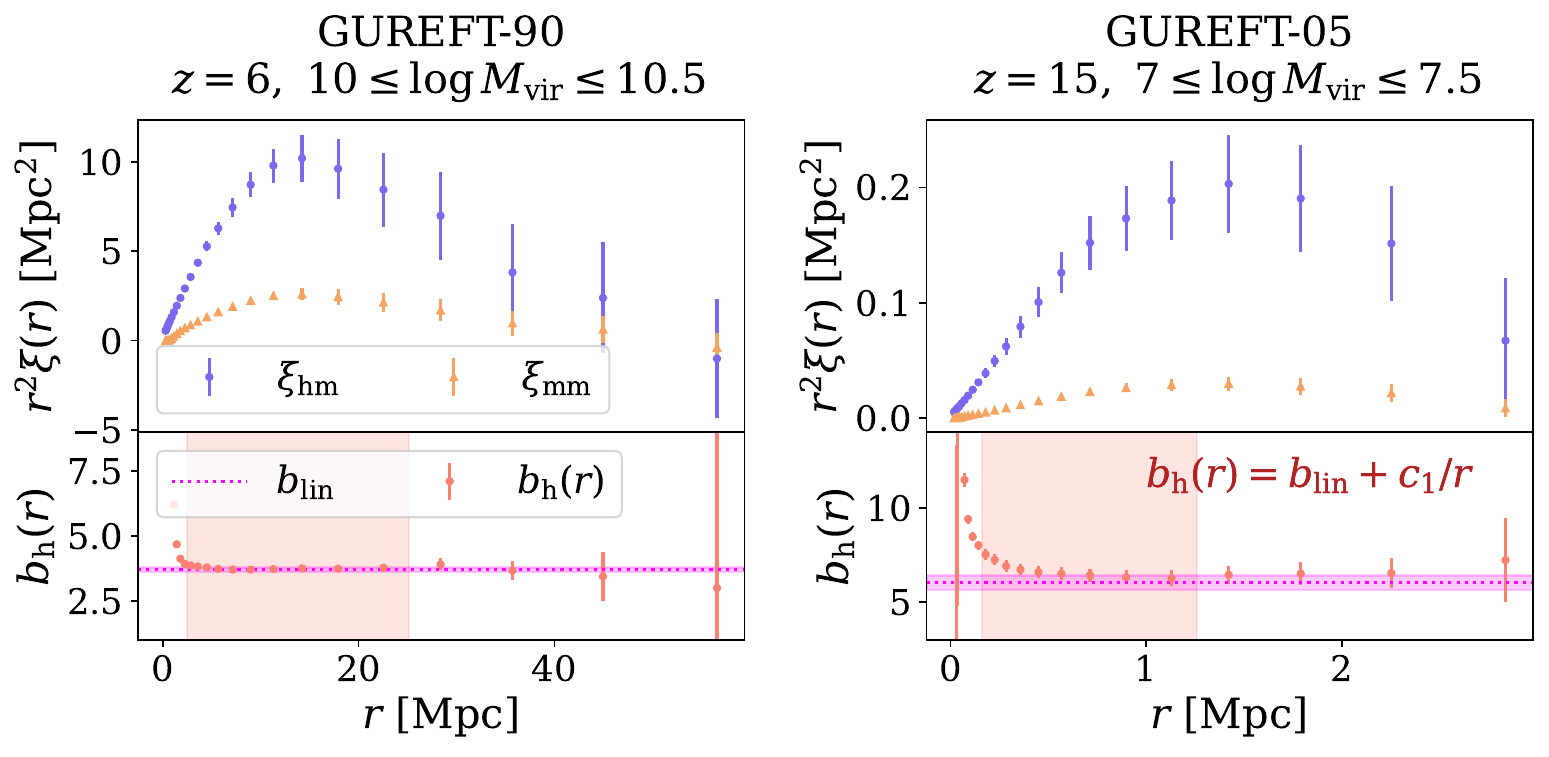}
    \caption{Two representative examples of our correlation function measurements and halo bias determination.
    The top panels show the halo–matter cross-correlation function and the matter auto-correlation function plotted as $r^2\xi(r)$.
    The bottom panels show the corresponding scale-dependent halo bias $b_{\rm h}(r)$.
    The error bars show jackknife uncertainties.
    The vertical shaded regions indicate the scales used for fitting \autoref{eq:bh_vs_blinear}, whereas the horizontal dotted lines and shaded regions show the best-fits and uncertainties of the linear bias $b_{\rm lin}$.}
    \label{fig:cf}
\end{figure*}

The top panels of \autoref{fig:cf} show the correlation function measurements for two representative samples, one high-mass, low-redshift, and one low-mass, high-redshift.
The halo–matter cross-correlation $\xi_{\rm hm}$ exceeds the matter auto-correlation $\xi_{\rm mm}$, indicating that halos are more strongly clustered than the underlying matter field.
Here, and throughout, we estimate jackknife uncertainties by dividing each box into $4^3$ cubical cells of identical volume, and leaving out one box at a time to calculate the correlation function and bias for the rest of the box.
We then estimate the uncertainty from the resulting covariance between realizations.
We find agreement with other jackknife schemes to better than $\lesssim 10\%$, which suffices for our purposes.
As expected, $\xi_{mm}$ has smaller uncertainties due to the larger number of dark matter particles.
We make our measurements publicly available at \url{https://github.com/KuanWang-Astro/GUREFT_CF/tree/main}.

\subsection{Halo bias estimation}
\label{sec:bias_method}

For a given halo sample, the bias $b_{\rm h}$ can be measured from two-point correlation functions, with
\begin{equation}
\label{eq:bhm}
   b_{\rm h} = \xi_{\rm hm} / \xi_{\rm mm},
\end{equation}
or $b_{\rm h}^{(\rm auto)} = (\xi_{\rm hh} / \xi_{\rm mm})^{1/2}.$ 
We will present bias results from cross-correlations, which are less sensitive to shot noise, and yield more reliable estimates, especially in the case of small sample sizes.
However, we have checked that the auto-correlation approach yields results that qualitatively agree with our main results.

In this first study we aim to measure the scale-independent linear halo bias $b_{\rm lin}$.
However, the $b_{\rm h}$ we measure from correlation functions can be scale dependent, i.e., $b_{\rm h}=b_{\rm h}(r)$.
At small scales, nonlinear gravitational effects, halo exclusion, mode coupling, and environmental dependences can all modify $b_{\rm h}(r)$ \cite[see Ref.][and references therein]{desjacques18}.
Ideally, at large scales, $b_{\rm h}(r)$ approaches the constant linear bias.
Practically, however, because of the limited box sizes that are available, the scale-independent regime is only partially sampled.
Also, the absence of long-wavelength modes in periodic simulations suppresses large-scale variance, biasing measurements near the box scale.

To mitigate these issues, we estimate the linear halo bias $b_{\rm lin}$ by performing a fit to the measured halo bias
\begin{equation}
    b_{\rm h}(r) = b_{\rm lin} + c_1/r,
    \label{eq:bh_vs_blinear}
\end{equation}
where $c_1$ is a free parameter, and the $1/r$ term is chosen to provide a minimal description of the dominant correction that decays toward zero at large scales, analogous to the scale‑dependent fits discussed in e.g., Ref.~\cite{smith07}.
In order to avoid issues near the halo and box boundaries, we only fit between approximately 30 times the virial radius of the typical halo in the sample and 1/6 the box size.
This process is illustrated in the bottom panels of \autoref{fig:cf}.
For each sample, we select and fit the intermediate range (vertical shaded region) of scales of $b_{\rm h}(r)$ for $b_{\rm lin}$ (horizontal line), again obtaining an uncertainty estimate (horizontal band) from jackknifing.
The jackknife uncertainties in $b_{\rm h}$ are smaller than those of the individual correlation functions, owing to partial cancellation of fluctuations when forming their ratio.
We test and confirm that the fitted linear bias value is not sensitive to reasonable changes in the choice of this range.

\subsection{Effective peak height}
\label{sec:nu_eff}

It is customary to describe the linear halo bias as a function of the halo peak height,
$\nu=\delta_c/\sigma(M, z)$,
where $\delta_c=1.686$ is the critical overdensity required for collapse and $\sigma(M, z)$ is the linear-theory rms matter fluctuation on mass scale $M$ at redshift $z$.
The halo peak height quantifies the rarity of a halo, and is dependent on the underlying cosmology.
We use \texttt{colossus} \cite{diemer2018} for the peak height calculation.
To account for the finite simulation volume, we truncate the linear power spectrum below $k_{\rm box} = 2\pi / L_{\rm box}$ when computing $\sigma(M, z)$, removing contributions from  the absent long-wavelength modes \cite[e.g.,][]{bagla06}.
We find that applying this correction brings into agreement the results from different box sizes.
We define the effective peak height $\nu_{\rm eff}$ for a halo sample as the mean peak height of all halos in the sample.

\section{Results}
\label{sec:results}

\subsection{Bias measurements}
\label{sec:bias_measure}

\begin{figure}
    \centering
    \includegraphics[width=\linewidth]{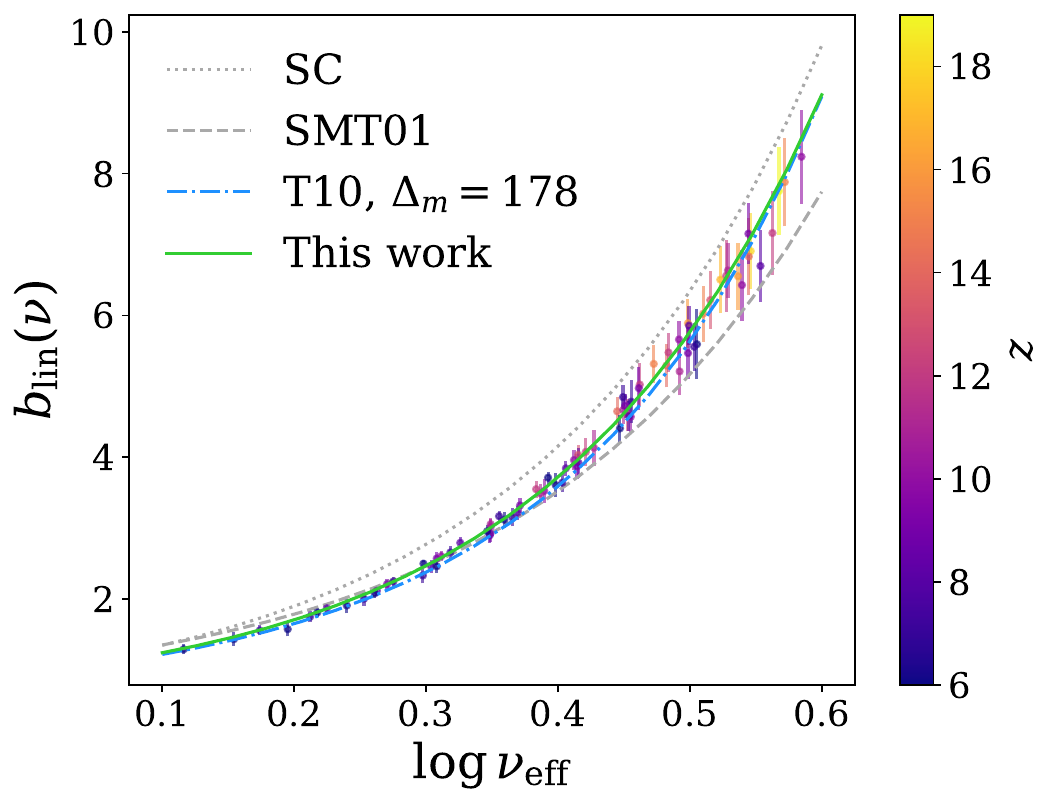}
    \caption{Measured halo bias–peak height relation and comparison to previous fits.
    The error bars show the measured halo bias against the effective peak height for each sample, with jackknife error, and are color coded by redshift.
    The different model fits, including our updated fit, are shown in curves of different colors and line styles, as labeled in the figure.
    The T10 fit was calibrated at $z\sim 0$ - 2.5 for $-0.4\lesssim\log_{10}\nu_{\rm eff} \lesssim 0.55$.}
    \label{fig:comp_fit}
\end{figure}

In \autoref{fig:cf}, we have shown the measurement process of halo bias.
For each mass bin sample of halos, we estimate the linear halo bias from correlation functions, and calculate the effect peak height.
We present measurements for all the available samples in \autoref{fig:comp_fit}, in terms of halo bias as a function of peak height, where each colored point represents one of our samples.
Different colors correspond to measurements at different redshifts, and the error bars represent jackknife uncertainties.
Higher peak heights corresponds to rarer peaks, which are more biased.
Our samples probe a peak height range of $1.31\leq \nu\leq3.88$ ($0.12\leq \log_{10} \nu\leq0.59$).
This range spans from low-mass halos below the typical galaxy-hosting scale to rarer systems that host galaxies on the bright end of the UV luminosity function.
Within our redshift range, we do not observe a clear dependence of the $b_{\rm lin}–\nu$ relation on redshift, apart from the implicit shift toward higher $\nu$ values at higher redshift.

\subsection{Comparison to existing fits}
\label{sec:fits_comp}

We compare our measurements in \autoref{fig:comp_fit} against several commonly adopted fitting functions in the literature.
In particular, we compare against the fitting formulae from:
\begin{itemize}
    \item the predictions of spherical collapse \cite[][hereafter SC]{cole-kaiser89,mo-white96}:
    \begin{equation}
        b_{\rm lin}(\nu) = 1+\frac{\nu^2-1}{\delta_c}
    \end{equation}
    \item the fit from~\citet{smt2001}(hereafter SMT01), motivated by the ellipsoidal collapse model and calibrated using simulations:
    \begin{equation}
    \begin{split}
    b_{\rm lin}(\nu) = 1 + \frac{1}{\sqrt{a}\,\delta_c} \Bigl[\, 
    & \sqrt{a}\,(a\nu^2) + \sqrt{a}\,b\,(a\nu^2)^{1-c} \\
    & - \frac{(a\nu^2)^c}{(a\nu^2)^c + b\,(1-c)(1-c/2)} \,\Bigr],
    \end{split}
    \end{equation}
    with $a=0.707,b=0.5,c=0.6$.
    \item the fit in~\citet{tinker2010} (hereafter T10, see also Ref.~\cite{mcclintock19}), also calibrated against simulations, with a more flexible form:
    \begin{equation}
    b_{\rm lin}(\nu)=1-A\frac{\nu^a}{\nu^a+\delta_c^a}+B\nu^b+C\nu^c,
    \label{eq:t10}
    \end{equation}
    where the parameters ($A,B,C,a,b,c$) are dependent on the virial overdensity, $\Delta$, with respect to the mean density of the Universe, as detailed in Table 2 of T10.
    Across our redshift range, the value of $\Delta$ is close to 178, so we treat it as a constant in our analysis.
\end{itemize}

\autoref{fig:comp_fit} overplots the predictions of these fitting formulae as curves.
Our measurements are clearly lower than SC and higher than SMT01, highlighting how these lower-$z$ calibrated fits do not apply to early times.
We find broad agreement with the T10 fit, which was calibrated against simulations at $z\sim 0-2.5$ (though at $-0.4\lesssim\log_{10}\nu_{\rm eff} \lesssim 0.55$, lower than our samples).
Interestingly, we find that adopting $\delta_c=1.524$ instead of 1.686 in the SC prediction provides a reasonable description of our measurements, but this value yields a prediction that is significantly below T10 in the low-$\nu$ regime.

\begin{figure*}[bt]
    \centering
    \includegraphics[width=0.8\linewidth]{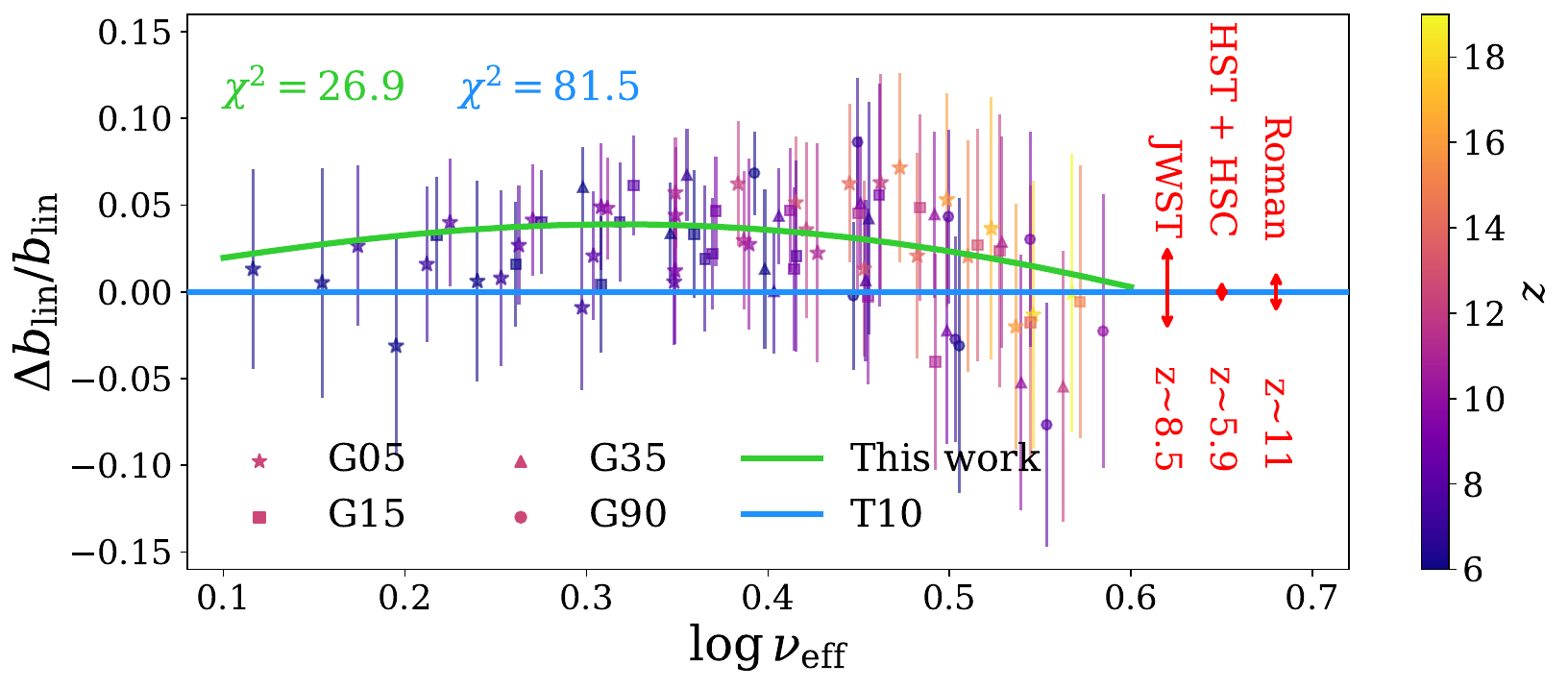}
    \caption{Comparison between measurements and fits.
    Same as \autoref{fig:comp_fit}, but in terms of fractional deviations from the T10 predictions.
    The $\chi^2$ values in the figure correspond to our updated fit and the original T10 fit, respectively.
    The arrows on the right summarize representative claimed measurements and forecasts of the fractional uncertainty in galaxy bias at high redshift from various facilities: a measurement at $z\sim8.5$ from Ref.~\cite[][JWST]{dalmasso26}, a measurement at $z\sim5.9$ from Ref.~\cite[][HSC]{harikane22}, and a forecast at $z\sim11$ from Ref.~\cite[][Roman]{munoz23}.}
    \label{fig:new_fit}
\end{figure*}

\autoref{fig:new_fit} compares our bias measurements against the T10 fit more closely, where it is clear that the T10 fit underpredicts the bias by a small but persistent amount, reaching a $\approx -4\%$ offset.
In the next subsection, we seek an improved fit within the T10 framework.

\subsection{Updated fit}
\label{sec:new_fit}

In the T10 formula, the parameters $\{A, B, C, a, b, c\}$ depend on redshift only through $\Delta$.
However, in the redshift range we consider, $\Delta$ is effectively constant, and $A\simeq1$.
Therefore, we allow $B, C, a, b, c$ to vary as free parameters with no redshift dependence, and fit them to our data points.
Following T10, we require that the parameters obey the relation 
\begin{equation}
    \int b(\nu)f(\nu){\rm d}\ln\nu=1.
\end{equation}
We adopt the halo mass function $f(\nu)$ from Ref.~\cite{yung25}, which is calibrated to describe \textsc{GUREFT} results.
In practice, we enforce that 
\begin{equation}
    \int b_{\rm new}(\nu)f(\nu){\rm d}\log_{10}\nu=\int b_{\rm T10}(\nu)f(\nu){\rm d}\log_{10}\nu
\end{equation}
within 1\% error.

Since our jackknife covariance matrix is rank-deficient, we perform the fit using diagonal errors only; for our mildly nonlinear model, we do not expect the best-fit parameters to be significantly affected by this choice.
The best fit we find is $a=0.08059, b=c=2.138$, i.e., the third and fourth terms in \autoref{eq:t10} have the same index, and $B+C=0.4499$, so the formula reads:
\begin{equation}
    b_{\rm lin}=1-\frac{\nu^{0.08059}}{\nu^{0.08059}+\delta_c^{0.08059}}+0.4499\nu^{2.138}.
    \label{eq:biasfit_us}
\end{equation}
This is our main result.
Assuming diagonal covariance matrix, these updated parameters reduce the $\chi^2$ value from 81.5 for the original T10 fit to 26.9, and the mean systematic offset from the measurements becomes $\sim-0.7\%$.
We do not quote $\chi^2/{\rm d.o.f.}$ values as the correlated data points render this statistic less meaningful.
We plot the updated fit in \autoref{fig:comp_fit} and \autoref{fig:new_fit}, in comparison to the original T10 formula.

We note that we do not perform an exhaustive search of the parameter space, and only present a possible improved fit that better describes the new measurements.
Our fit applies to the redshift and peak height range covered by the \textsc{GUREFT} samples, i.e., $6\leq z\leq19$ and $0.12\leq \log_{10}\nu\leq0.59$.
Within this range, the discrepancy with T10 is most pronounced at intermediate peak heights.
At the lowest peak height explored by T10, $\log_{10}\nu\simeq -0.4$, our extrapolated fit is slightly lower than the T10 fitting function by approximately 4\%.
However, the measurements quoted in T10 are also slightly lower than their fit in this region.
It is in principle possible to perform a joint fit to the T10 measurements and our simulation data to obtain a unified model across a wider range in peak height. 
However, given that this regime is not relevant for the high-redshift halos of interest here, we leave such an extension for future work.
Our fit also falls below the T10 fitting function when extrapolated to high $\nu$ values beyond the range probed by both our data and T10, and the bias behavior of these extremely rare halos remains an open question requiring further investigation.
We suggest that the updated fit in \autoref{eq:biasfit_us} be used for survey analyses at $z\gtrsim6$, where our simulations are calibrated, with a transition to the T10 solution as $z\to 0$.

\autoref{fig:new_fit} also places our results in observational context by showing several claimed measurements and forecasts of the fractional uncertainty in high-redshift galaxy bias from new and upcoming telescopes.
These error bars are comparable to --- or smaller than --- the halo bias discrepancy with T10, underscoring the significance of a sub-percent bias measurements, as provided in this work, when interpreting galaxy clustering observations.

\section{Discussion}
\label{sec:discussion}

Past work found no consensus on halo bias measurements at high redshifts. 
For instance, Ref.~\cite{cohn08} found $b(\nu)$ close to spherical-collapse predictions at $z=10$, systematically higher than T10, while Ref.~\cite{wang22} reported lower values than all existing fits at $z=5.5,7.7$, and 9.9.
Differences in simulation methods, halo definition and selection, and analysis techniques likely account for these discrepancies.
Specifically, insufficient numerical resolution can systematically shift halo masses, thereby altering the inferred peak height and bias, while finite-volume effects can suppress long-wavelength modes and bias clustering amplitudes low.
The choice of halo finder also introduces significant variance: friends-of-friends algorithms tend to link nearby structures and assign higher masses \cite{lukic09}, generally yielding a higher bias at fixed abundance, whereas spherical overdensity definitions depend sensitively on the chosen threshold.
Finally, analysis choices, such as the estimator employed (e.g., $\xi_{hm}$ vs.\ $\xi_{hh}$) and the scale range over which the bias is fit, can introduce systematic offsets, as both including quasi-linear scales and shot noise from auto-correlations generally elevate the inferred bias.
In this work, we have overcome these issues by using the tailored simulations with phase-space halo finding, applying corrections for finite volume effects, and selecting the appropriate scale range for linear bias fitting.

Looking ahead, improved simulations will be critical for refining halo bias predictions at high redshift.
Larger volumes are needed to reduce cosmic variance and enable robust measurements of rare, massive halos \cite[see, e.g., Ref.][for a study of massive quasar-hosting halos at $z\simeq4$]{pizzati24}, while higher resolution is required for objects below our lowest mass of $M_h\sim 10^{6.5}\, M_\odot$. 
Simulations spanning alternative cosmologies will test the robustness of the $b(\nu)$ relation and determine whether the offset we find persists across models.
Targeted studies of extreme environments such as protoclusters and voids can further clarify environmental effects not captured in typical volumes.
Several simulation efforts \cite[e.g.,][]{ocvirk2016,xu2016,wang22,hernandez-aguayo2023} have addressed some of these aspects, yet runs designed for high-$z$ analyses are necessary to robustly model halo clustering at early times.

In this first work we have assumed that halo bias is ``universal'' (i.e., only depends on peak height), but clustering may vary with properties such as formation history, concentration, and environment, giving rise to ``assembly bias'' \cite{croton07,dalal08}.
This can lead to departures from a simple $b_{\rm lin}(\nu)$ relation \cite{lazeyras16,paranjape17,mansfield20}, which is not constrained by our measurements. 
Likewise, we have limited ourselves to linear bias in this work, but there is further information on structure formation in the non-linear bias terms~\cite{modi17,gil-marin2010,jose2016,hoffman17,schmittfull19,mons26}.
These terms may be most important for rare or massive halos \cite{manera10,lazeyras16}, and have recently been studied at high $z$ in Ref.~\cite{boone26}.

\section{Conclusions}
\label{sec:conclusions}

Here we present new measurements of halo bias at $6\leq z\leq19$ using \textsc{GUREFT}, a simulation suite specifically designed to capture the mass and redshift ranges relevant to the first galaxies.
At these high redshifts, halos hosting observable galaxies occupy a more extreme peak-height regime, which is less well constrained by low-redshift calibrations.
Our results provide a new handle on halo clustering in the first billion years, extending our understanding of large-scale structure and enabling precise interpretation of high-redshift galaxy surveys and line-intensity mapping data.
Given the measured or forecasted precision of these surveys, even small deviations in halo bias are observationally significant, both for interpreting galaxy clustering and for refining models of galaxy formation and cosmology.

We carefully account for systematic effects in determining the relation between halo bias and peak height.
Our results broadly agree with the fitting function provided by T10, which was calibrated at low redshift.
However, we find consistently higher biases for the peak heights probed.
At $\log_{10} \nu \simeq 0.2$ - 0.5, where our measurements have better precision than T10, the discrepancy is at the 3 - $4\%$ level.
We provide an updated fit to the T10 functional form for $b(\nu)$ at these high redshifts, given in \autoref{eq:biasfit_us}, which reduces the mean systematic offset from the measurements to $<1\%$.

In summary, high-redshift structure formation offers a new way to test cosmology and astrophysics that is within the reach of current observatories.
To take full advantage of observations, theory must match their precision.
Our work provides a first step toward that goal, delivering accurate measurements of linear halo bias during cosmic dawn and laying the groundwork for interpreting observations of the earliest galaxies.

\begin{acknowledgments}
\subsection*{Acknowledgements}
We wish to thank Mike Boylan-Kolchin and Jeremy Tinker for discussions and comments.
This work has been supported at UT Austin by the HETDEX Cosmology Fellowship, NSF Grants AST-2307354 and AST-2408637, and the NSF-Simons AI Institute for Cosmic Origins.
LYAY is supported by a Giacconi Fellowship from the Space Telescope Science Institute, which is operated by the Association of Universities for Research in Astronomy, Incorporated, under NASA contract HST NAS5-26555 and JWST NAS5-03127.

\end{acknowledgments}

\bibliographystyle{apsrev4-2}
\bibliography{main}

\end{document}